\documentstyle[prl,aps,epsf,floats,twocolumn]{revtex} 
\newcommand{\be}{\begin{equation}}
\newcommand{\ee}{\end{equation}}
\begin{document}
\draft
\twocolumn[\hsize\textwidth\columnwidth\hsize\csname
@twocolumnfalse\endcsname

\title{Fitness-dependent topological properties of the World Trade Web}
\author{Diego Garlaschelli$^{1,2}$, Maria I. Loffredo$^{2,3}$}
\address{$^1$Dipartimento di Fisica, Universit\`a di Siena, Via Roma 56, 53100 Siena ITALY\\
$^2$INFM UdR Siena, Via Roma 56, 53100 Siena ITALY\\
$^3$Dipartimento di Scienze Matematiche ed Informatiche, Universit\`a di Siena, Pian dei Mantellini 44, 53100 Siena ITALY}
\date{\today}
\maketitle
\begin{abstract}
Among the proposed network models, the hidden variable (or good get richer) one is particularly interesting, even if an explicit empirical test of its hypotheses has not yet been performed on a real network. Here we provide the first empirical test of this mechanism on the world trade web, the network defined by the trade relationships between world countries. We find that the power-law distributed gross domestic product can be successfully identified with the hidden variable (or fitness) determining the topology of the world trade web: all previously studied properties up to third-order correlation structure (degree distribution, degree correlations and hierarchy) are found to be in excellent agreement with the predictions of the model. The choice of the connection probability is such that all realizations of the network with the same degree sequence are equiprobable.
\end{abstract}

]
\narrowtext
Networks emerge almost ubiquitously in complex systems, from the cell to the Internet and the economy \cite{barabba,mendes,siam}. Despite differences in their nature, many real-world networks are characterized by similar topological properties, strikingly different from those displayed by simple random graphs. This finding has motivated the search for theoretical models aimed at understanding the mechanisms at the basis of network organization. 
Currently, such models can be grouped in two main classes: a first class exploiting the \emph{rich get richer} or \emph{preferential attachment} mechanism \cite{barabba}, where the topology evolves in such a way that already well connected vertices become more and more connected in a multiplicative fashion, and a second class based on the \emph{good get richer} \cite{fitness} or \emph{hidden variable} \cite{fitness,pastor} hypothesis, where vertices are assumed to be characterized by an intrinsic quantity or \emph{fitness} which determines their connection probability. The former mechanism is particularly suitable to model evolving networks whose future topology is likely to be determined essentially by their past one, without any additional `external' information. The latter mechanism is better adapted to model static networks where the topological properties are essentialy determined by some additional, `physical' information of non-topological nature but intrinsically related to the role played by each vertex in the network.

With suitable implementations, both types of models were shown to successfully reproduce a range of nontrivial topological properties\cite{barabba,mendes,fitness,pastor}. However, the main question remains whether the basic hypotheses of the models can be tested satisfactorily in each particular case study, so that one can decide which theoretical mechanism best captures the empirical organizing principle of the specific system. While the effective presence of  preferential attachment has been detected in some evolving networks such as the Internet, collaboration and coauthorship networks \cite{attach1,attach2}, no analogous test for the hidden variable mechanism has been so far performed on those networks which are the natural candidates for it.
Indeed, in most cases the hidden variable is regarded as a hypothetical quantity with no clear physical interpretation \cite{pastor} or as a statistical one equal to the \emph{desired degree} (or \emph{fugacity}) \cite{newman} of a vertex. 
In the present Letter, we provide complete evidence that the hidden variable mechanism correctly reproduces all the relevant topological properties of a real network: the \emph{World Trade Web} (WTW in the following), or the network formed by the trade relationships between all world countries \cite{wtw}. 
Our main result is that the \emph{Gross Domestic Product} (GDP in the following) can be identified with the \emph{fitness} variable that, once a form of the connection probability is introduced, completely specifies the expected topological properties of the WTW. Remarkably, all expected trends are in excellent agreement with the empirically observed ones.

The properties of the WTW corresponding to the year 2000 were already studied in ref. \cite{wtw} and reveal a complex organization of international trade channels. Moreover, the degree (number of trade partners) of a country was found to be correlated with its \emph{per capita} GDP, even if a number of exceptions to this trend were found \cite{wtw}.
Here we present a finer analysis based on a different, more detailed data set \cite{data} which reports \emph{all} trade relationships for each country, and not only a limited number of them as in ref. \cite{wtw}. This allows us to extend the previous analysis to a wider region, with interesting results.
Moreover, besides the values of \emph{per capita} GDP, the present data set also reports the population sizes of all world countries, so that the corresponding \emph{total} GDP (which we denote by $w_i$ for each country $i$) can be easily obtained. Since we expect that the total GDP is more closely related to the trade activity of a country than the \emph{per capita} GDP is, the knowledge of $w_i$ allows us to consider its value as the natural candidate to be identified with the \emph{hidden variable} associated to each country $i$ in the WTW. 
In other words, since the WTW is defined by the exchange of wealth between its vertices, we identify the fitness (potential ability of being connected to other vertices \cite{fitness}) of a vertex with its wealth, which here is measured by the GDP of the corresponding country. In order to have an adimensional quantity we define the \emph{fitness} $x_i$ of a country $i$ as its relative GDP:
\be\label{x}
x_i\equiv\frac{w_i}{\sum_{j=1}^Nw_j}
\ee
where clearly $0<x_i<1\quad\forall i$. In the following we will show the results of the analysis of the WTW and GDP for the year 1995, even if the data set covers a longer time interval \cite{data} and we analysed several different years obtaining analogous results. For the 1995 data, the number of countries equals $N=191$ and the number of links is $L=16,255$. Note that we are referring to the \emph{undirected} version of the WTW (also studied in ref.\cite{wtw}), where two countries $i$ and $j$ are connected if a nonzero trade flow exists in any direction between them ($i$ exports to $j$, $j$ exports to $i$ or both conditions apply). The reason for this choice is that, as we show below, a suitable form of the connection probability can be introduced which correctly reproduces the undirected version of the network and at the same time has an interesting economic meaning.

\begin{figure}[ht]	
\centerline{\epsfxsize 3.5 truein \epsfbox{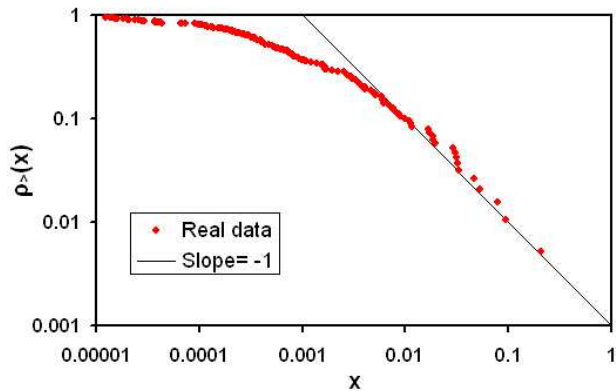}}
\caption[]{
\label{fig_rho}
\small Cumulative fitness distribution (circles) of world countries and power law with slope -1 (solid line).}
\end{figure}

In the \emph{fitness model} \cite{fitness}, once the set of fitness values $\{x_i\}_{i=1}^N$ is fixed, the expected topological properties of the network only depend on the statistical distribution $\rho(x)$ of the fitness and on the probability $f(x_i,x_j)$ that the vertices $i$ and $j$ are connected. 
The cumulative fitness distribution $\rho_>(x)\equiv\int_x^\infty \rho(x')dx'$ for our system is shown in fig.\ref{fig_rho} and is found to display a power-tail of the form $\rho_>(x)\propto x^{1-\tau}$, corresponding to a tail in the probability density of the form $\rho(x)\propto x^{-\tau}$. A power-law with exponent $1-\tau=-1$ (corresponding to $\tau=2$) is shown as a reference for the large $x$ range. This \emph{Pareto tail} is common to many other  wealth distributions \cite{pareto}.

Now we must specify a form of the connection probability. Consistently with our expectation that the GDP of a country represents its potential ability to develop trade relationships with other countries, we interpret the GDP (up to a proportionality constant) as the \emph{desired degree} of a country in the same spirit of ref.\cite{newman}. Then, using the same arguments of the cited reference to require that no two vertices have more than one edge connecting them, we end up with the following form of the connection probability:
\be\label{f}
f(x_i,x_j)=\frac{\delta x_ix_j}{1+\delta x_ix_j}
\ee
where $\delta>0$ is the only free parameter of the model. 
The idea leading to the above choice was first proposed in ref. \cite{maslov} to detect nontrivial patterns of a real network by comparing it with a suitably randomized one with the same degree sequence. This procedure also allows to check if some properties of the network are merely due to the impossibility of having multiple links \cite{newman,maslov}.
Once the dependence on $x$ is made explicit, the requirement of ref.\cite{maslov} translates to eq. (\ref{f}). With such a choice, once the set of fitness values $\{x_i\}_{i=1}^N$ is specified, all realizations of the network with the same degree sequence $\{k_i\}_{i=1}^N$ occur in the ensemble of all possible ones with the same probability \cite{newman,maslov}. 
In other words, all the networks obtained through permutations of the links that leave the degree sequence unchanged are equiprobable.
From an economic point of view, this means that once the GDP of world countries is specified, the probability of having a particular realization of the WTW only depends on the degree sequence, which here is the vector of the numbers of trade partners.

The form of the connection probability $f(x_i,x_j)$ completely specifies the topological properties of the network. For instance, one \emph{zero}th-\emph{order} property is the expected number of links, which is given by:
\be\label{L}
\tilde{L}=\frac{1}{2}\sum_{i=1}^N\sum_{j\ne i}f(x_i,x_j)
\ee
and determines the expected mean degree $\langle\tilde{k}\rangle=2\tilde{L}/N$ of the network (throughout the Letter, the expected value of a quantity will be denoted by a tilde). This can be used in order to tune the parameter $\delta$ to a value such that $\tilde{L}$ equals the actual number $L$ of links in the network. We find that this is obtained when $\delta=80 N^2$, and all the following results of the present Letter correspond to this single choice of the parameter. 
Note that in eq.(\ref{L}) and in what follows we avoid the use of integrals since the discrete sums can be directly computed on the empirical values of $x_i$ without introducing an approximated analyitical form for $\rho(x)$ which would result in loss of information. In this way we completely focus on the process yielding the expected quantities once the values $\{x_i\}_{i=1}^N$ are given as an input.

\begin{figure}[ht]	
\centerline{\epsfxsize 3.5 truein \epsfbox{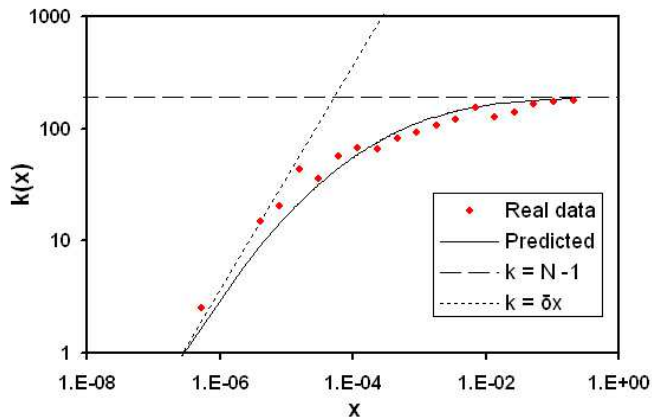}}
\caption[]{
\label{fig_k(x)}
\small Comparison of the observed and expected dependence of the degree $k$ on the fitness $x$ (the circles are obtained through a logarithmic binning of raw data). The asymptotic trends $\tilde{k}=N-1$ and $\tilde{k}\approx \delta x$ are also shown.}
\end{figure}

A \emph{first-order} property that can be used to test the predictions of the model is the dependence of the expected degree $\tilde{k}_i$ of each vertex $i$
\be\label{k}
\tilde{k}_i=\sum_{j\ne i}f(x_i,x_j)
\ee
on the fitness $x_i$. In fig.\ref{fig_k(x)} we compare the predicted behaviour of $\tilde{k}_i(x_i)$ (solid line) with the empirical one of $k_i(x_i)$ (circles) and find a close agreement between them. As expected, richer countries have more connections than poor ones. More precisely, $\tilde{k_i}$ is an increasing function of $x_i$ such that $\tilde{k}_i\to\ N-1$ when $x_i\to\infty$. This corresponds to the requirement that no two edges are connected by more than one link, and closely reflects the results of ref.\cite{newman}. In the opposite limit $x_i\approx 0$ one has $f(x_i,x_j)\approx \delta x_i x_j$ and therefore ${\tilde{k_i}\approx\delta x_i\sum_j x_j\approx\delta x_i}$ (both asymptotic trends are shown as a dashed and a dotted line respectively).
Let us stress that the predicted values $\tilde{k}_i$ are obtained only through the empirical GDP data without any additional information about the topology of the network. By contrast, the actual values $k_i$ are obtained only from the WTW data, which is an independent source of information. This remarkable comment extends to every result that will be presented below.

\begin{figure}[ht]	
\centerline{\epsfxsize 3.5 truein \epsfbox{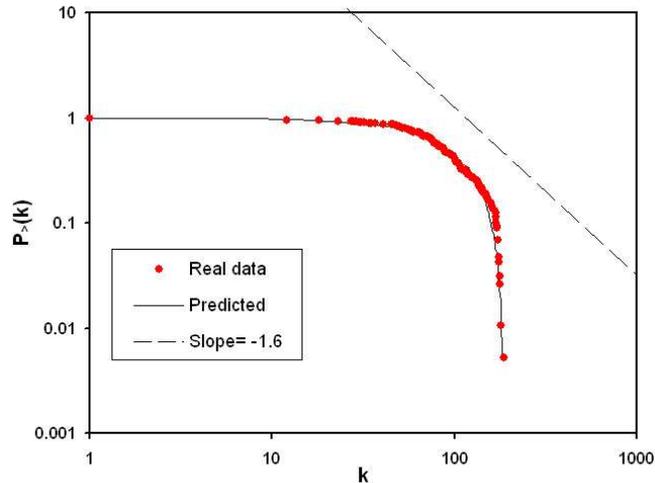}}
\caption[]{
\label{fig_p(k)}
\small Cumulative degree distributions $P_>(k)$: empirical (circles) and predicted (solid line). A power law with slope $-1.6$ is also shown (dashed line).}
\end{figure}

Once the expected values $\tilde{k}_i$ are given, one can plot the corresponding statistical distribution. In fig.\ref{fig_p(k)} we compare the empirical cumulative degree distribution (circles) with the predicted one (solid line), and we find that they are in excellent agreement.
As a comparison, note that the empirical degree distribution presented in ref.\cite{wtw} extends to a narrower region and diplays a less pronounced cut-off for large $k$. This is probably due to the fact that in that case the data set reports only the 40 most relevant links for each country, even if by a symmetry argument the authors recover many (but not all) missing observations. This leads the authors to fit the data with a power law $P_>(k)\propto k^{1-\gamma}$ with $1-\gamma=-1.6\pm 0.1$ (shown as a reference in fig.\ref{fig_p(k)}). Here instead we find that the power-law region is only a small part of the whole degree distribution and we conclude that the WTW is \emph{not} a scale-free network. The sharp cut-off for large $k$ corresponds to the saturation effect $k(x)\to N-1$ for large $x$ shown in fig.\ref{fig_k(x)}, and it is in accordance with the theoretical results of ref.\cite{newman}. 

\begin{figure}[ht]	
\centerline{\epsfxsize 3.5 truein \epsfbox{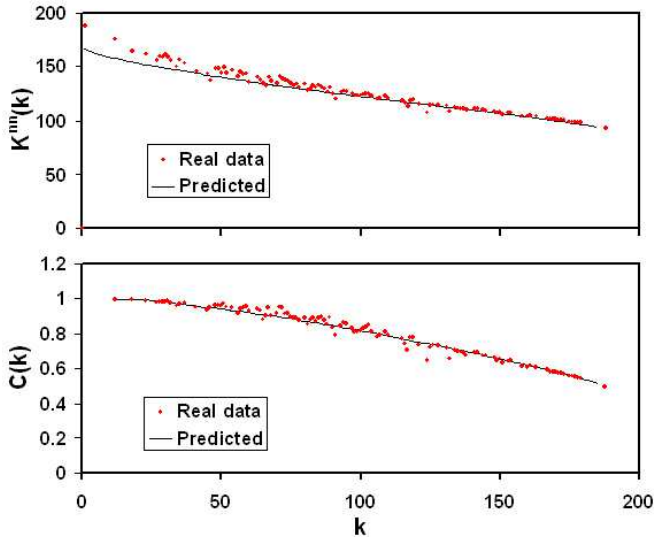}}
\caption[]{
\label{fig_both}
\small Average nearest neighbour degree $K^{nn}$ and clustering coefficient $C$ versus the degree $k$ for the real (circles) and model (solid line) webs.}
\end{figure}

In ref.\cite{wtw} it was shown that the WTW is also characterized by higher-order properties which require the knowledge of correlations between vertex degrees. For instance, the \emph{average nearest neighbour degree} $K^{nn}(k)$ (defined as the mean degree $K^{nn}_i$ of the neighbours of a vertex $i$, averaged over all vertices with the same degree $k_i$) is a \emph{second-order} topological property requiring the knowledge of the two-vertices conditional probability $P(k'|k)$ that a vertex with degree $k$ is connected to a vertex with degree $k'$. The value of $K^{nn}$ predicted by the fitness model \cite{fitness,pastor} is
\be\label{ANND}
\tilde{K}^{nn}_i=\frac{\sum_{j\ne i}\sum_{k\ne j}f(x_i,x_j)f(x_j,x_k)}{\tilde{k}_i}
\ee
In fig.\ref{fig_both} we plot the values of $K^{nn}_i$ versus $k_i$ for the empirical network and compare them with the predicted ones $\tilde{K}^{nn}_i$. The accordance is again very good, apart from a slight underestimation of $K^{nn}$ for small $k$. 
The decreasing trend of $K^{nn}(k)$ signals the presence of \emph{disassortativity}, or anticorrelation between vertex degrees \cite{assort,internet}. Poorly connected countries are on average connected to well connected (richer) ones. By contrast, in an uncorrelated network $K^{nn}$ is constant. Remarkably, while in ref.\cite{newman} it was shown that the mechanism suggested in ref.\cite{maslov} and based on the connection probability of eq.(\ref{f}) explained only some of the observed disassortativity of the Internet, here the model succeeds in providing a complete description of the observed correlation structure.
It should be stressed that, despite a power-law fit of the form $K^{nn}(k)\propto k^{-\nu}$ is commonly proposed \cite{wtw,internet}, extensive numerical simulations \cite{newman} showed that the functional form of $K^{nn}(k)$ generated by the mechanism discussed here is different, even if in the large $k$ range it appears to approach a power-law behaviour. Consistently with such results, the WTW of ref.\cite{wtw} only displayed a power-law behaviour in a very narrow range. For this reason we avoid to fit our data with a power-law, and we prefer to plot them in linear scale as in fig.\ref{fig_both}. 

Finally, we focus on the \emph{clustering coefficient} $C_i$, which is a \emph{third-order} topological property defined as the connected fraction of neighbour pairs of a vertex $i$. In principle, the evaluation of this quantity requires the knowledge of the three-vertices conditional probability $P(k',k''|k)$ that a vertex with degree $k$ is simultaneously connected to two vertices
with degrees $k'$ and $k''$ \cite{pastor}. However, the fitness model is completely specified by the pairwise connection probability $f(x_i,x_j)$, and as a result the quantity $P(k',k''|k)$ can be simply expressed as $P(k',k''|k)=P(k'|k)P(k''|k)$ \cite{pastor}. The expected clustering coefficient $\tilde{C}_i$ is given by
\be\label{C}
\tilde{C}_i=\frac{\sum_{j\ne i}\sum_{k\ne j,i}
f(x_i,x_j)f(x_j,x_k)f(x_k,x_i)}{\tilde{k}_i[\tilde{k}_i-1]}
\ee
Figure \ref{fig_both} shows the dependence of the clustering coefficient on the degree (no power-law fit is provided for the same reason as above). The agreement between the predicted and the observed behaviour is again excellent. This strongly suggests that the real WTW is well approximated by a network with only  second-order nontrivial correlations, and that the third-order structure can be obtained in terms of the first- and second-order one. As noted in ref.\cite{wtw}, the decrease of $C(k)$ points out the presence of \emph{hierarchy} in the network \cite{hierarchy}, or the tendency of low-degree countries to trade with tightly interacting partners, and that of higher-degree countries to trade with more loosely interacting partners.

We stress again the very intriguing result that all the predicted properties of the WTW shown in this Letter are obtained by exploiting only the set of GDP values, and reproduce almost perfectly the actual properties computed on the independent WTW data. By providing a good average description of the global trade activity, our analysis also provides a basis for the characterization of the \emph{deviations} from this average behaviour, such as the detection of geographical preferences due to reduced transportation costs and the identification of countries with more (or less) trade partners than the mean expected value. We believe that the study of such deviations is an important point to be addressed in the future.
In regards to the more general problem of network modelling, our results strongly support the hypothesis of the \emph{hidden variable} or \emph{fitness} model \cite{fitness}, which is likely to capture the topological organization of many other real networks. The possibility of identifying the fitness variable with a physical quantity is a significant step forward, since in previous studies its distribution was chosen quite arbitrarily in order to yield the observed degree distribution of the network in an \emph{ad hoc} fashion. Instead, once the fitness distribution is fixed by observation as in the present case, the topological properties only depend on the functional form of the connection probability and the choice of the latter allows a deeper physical understanding of the network formation process.


\begin{thebibliography}{99}

\bibitem{barabba}
R. Albert and A.-L. Barab\'asi, \emph{Rev. Mod. Phys.} \textbf{74}, 47 (2002).

\bibitem{mendes}
S.N. Dorogovtsev and J.F.F. Mendes, \emph{Adv. Phys.} \textbf{51}, 1079 (2002).

\bibitem{siam}
M.E.J. Newman, \emph{SIAM Review} \textbf{45}, 167 (2003).

\bibitem{fitness}
G. Caldarelli, A. Capocci, P. De Los Rios, and M.A. Mu\~noz,
\emph{Phys. Rev. Lett.} \textbf{89}, 258702 (2002).

\bibitem{pastor}
M. Bogu\~n\'a and R. Pastor-Satorras, \emph{Phys. Rev. E} \textbf{68},
036112 (2003).

\bibitem{attach1}
M.E.J. Newman, \emph{Phys. Rev. E} \textbf{64}, 025102 (2001).

\bibitem{attach2}
H. Jeong, Z. N\'eda and A.-L. Barab\'asi,
\emph{Europhys. Lett.} \textbf{61} (4), 567 (2003).

\bibitem{newman}
J. Park and M.E.J. Newman, \emph{Phys. Rev. E} \textbf{68}, 026112 (2003).

\bibitem{wtw}
Ma \'A. Serrano and M. Bogu\~n\'a, \emph{Phys. Rev. E} \textbf{68}, 015101(R) (2003).

\bibitem{data}
K.S. Gleditsch, \emph{Journal of Conflict Resolution} {\bf 46}, 712 (2002).

\bibitem{pareto}
M. Levy and S. Solomon, \emph{Physica A} \textbf{242}, 90-94 (1997).

\bibitem{maslov}
S. Maslov, K. Sneppen, and A. Zaliznyak, \emph{Physica A} \textbf{333}, 529-540 (2004).

\bibitem{assort}
M.E.J. Newman, \emph{Phys. Rev. Lett.} \textbf{89}, 208701 (2002).

\bibitem{internet}
R. Pastor-Satorras, A. V\'azquez, and A. Vespignani, \emph{Phys. Rev.
Lett.} \textbf{87}, 258701 (2001).

\bibitem{hierarchy}
E. Ravasz and A.-L. Barab\'asi, \emph{Phys. Rev. E} \textbf{67}, 026112 (2003).

\end{thebibliography}
\end{document}